\documentclass[twoside,10pt,a4paper]{newFNLstyle}
\usepackage{graphics}
\usepackage{cite}

\begin{document}

\volnumpagesyear{0}{0}{000--000}{2003}
\dates{received date}{revised date}{accepted date}

\title{ONSET OF SHEAR WAVES IN A BACTERIAL BATH: A NOVEL EFFECT}
\authorsone{SUPURNA SINHA}
\affiliationone{}
\mailingone{Raman Research Institute,
Bangalore 560080,India.}

\maketitle

\markboth{Onset of Shear Waves in a Bacterial Bath: A Novel Effect}{Sinha}

\pagestyle{myheadings}
% Comment this out to remove the running heads


\begin{abstract}
Recent experiments on particle diffusion in bacterial baths indicate
the formation of correlated structures in the form of bacterial swirls.
Here we predict that such a structural ordering 
would give rise to the new effect of  
propagating
shear waves in a bacterial bath at length scales of the order
of a swirl, which corresponds to 
time scales of the order of the lifetime of a swirl.
Our prediction can be tested against future experiments in 
bacterial baths.  
\end{abstract}

\section{Introduction}
A simple liquid cannot sustain shear stress under a static condition.
However, at molecular length and time scales (i.e. time scales
of the order of the relaxation time of the liquid), it is known that a liquid
behaves much like a solid with a viscoelastic character \cite{hansen}.
One can understand such a phenomenon qualitatively as follows. On
molecular length
scales a dense liquid develops solid-like structures due to slowing down
of diffusional structural relaxation and consequently it can support a
shear wave.
Molecular dynamics simulations of hard sphere liquids and experiments on
liquid
Argon support this picture \cite{alley,trk}. On the theoretical end,
such a behavior can be understood
in terms of 
viscoelastic models\cite{hansen,trk,ssinha}.

In recent years there have been experiments probing `Brownian Motion' of
polysterene
spheres in bacterial baths \cite{libchaber}. These experiments point
to a crossover from superdiffusive to diffusive 
behavior of the mean-square
displacement of the polysterene beads suspended in a bacterial bath. 
This effect is similar to ballistic to diffusive transition in
colloids \cite{libchaber,cohen,schof} 
where the borderline for such a transition is given by 
the viscous damping time $t_0 = \frac{m_b}{\gamma}$
($m_b$ being the mass of the Brownian particle and $\gamma$ the
friction coefficient), which for a polysterene
bead of diameter $10 \mu m $ is about $10^{-5} sec$. However, in the case
of a bacterial bath one notices that such a transition takes 
place at a time of about $\tau_c\approx 1 sec $. This characteristic 
time scale $\tau_c$  
increases with
the number density $n$ of bacterial cells in the bath. The length scale
$l_c=\sqrt{[\Delta r^2(\tau_c)]}$, which is the root mean square 
displacement at $t=\tau_c$ also increases with $n$. These
observations point to the formation of correlated structures in the 
form of bacterial swirls which have been observed in these systems. In fact, one
finds that $l_c$ is indeed of the order of the spatial scales of these
swirls ($10-20 \mu m$). To sum up, one of the key observations coming out of
these experimental studies is the emergence of a time scale and a 
corresponding length scale in the Brownian motion of a polysterene bead
suspended in a bacterial bath.  
This time scale $\tau_c$ increases 
with the number density $n$ of
the bacterial bath, indicating the appearance of solid-like ordering over
length scales
around 
$l_c=\sqrt{[\Delta r^2(\tau_c)]}$. 
In a dense liquid
such ordering takes
place on molecular length scales. In the present experiments 
%the `liquid'under consideration is an active medium of biological origin. 
%Nonetheless, one can import
%some of the ideas from passive liquids to understand certain aspects of
%the dynamics
%of such a bath. In particular, in this paper we focus on the fact that
%such a structural
%ordering in the bacterial bath would indicate the appearance of a shear
%wave on
%length scales of the order of $l_c=\sqrt{[\Delta r^2(\tau)]}$. This is the
%focus of our
%present theoretical study which has experimental implications.
the bacterial bath can be viewed as a scaled up liquid with a typical
relaxation time scale $\approx 1 sec$ in contrast to a colloidal
liquid characterized by a time scale $\approx 10^{-5} sec$. 

\section{Viscoelasticity in Bacterial Bath}

We now describe in some more quantitative detail the crossover from 
super-diffusive to diffusive behavior in a bacterial bath. 
Consider the dynamics of a bead of mass $m_b$ suspended
in a bacterial bath \cite{libchaber} in terms 
of a phenomenological Langevin-type equation:
\begin{equation}
m_b\frac{d\vec{v}}{dt} = -\gamma \vec{v} +\vec{f}(t)
\label{lang}
\end{equation}
where $\gamma$ \cite{libchaber,cheung} is the 
friction coefficient of the fluid. 
%=6\pi \eta R$, $\eta$ being the viscosity of the fluid and
%$R$ the radius of the bead 
The collisional force $f(t)$ experienced by the 
bead due to ``kicks" from the swirls of bacteria has a zero mean and is 
exponentially correlated in time with a time scale $\tau_c$ corresponding to 
the lifetime of a swirl. Thus 
\begin{equation}
<\vec{f}(t)>=0;
<\vec{f}(0)\cdot\vec{f}(t)> = <{f^2(0)}> exp(-t/\tau_c).
\label{force}
\end{equation}
where $<{f^2(0)}> = D \gamma^2/\tau_c$ with $D=k_BT/\gamma$ the Einstein
diffusion coefficient. 
Such an analysis leads to the following behavior for the mean square
displacement of the bead. 
At short times ($t<<\tau_c$) the motion of the bead is ballisticlike 
with $<\Delta r^2(t)>\sim t^2$ whereas at long times ($t>>\tau_c$) 
it is diffusive
$<\Delta r^2(t)>\sim t$\cite{libchaber}. 

More recent experiments on correlated bacterial motion\cite{yhat} 
also indicate
the emergence of a long time scale $\tau_c$ increasing with the number
density
$n$ of bacteria. One interesting observation that comes out of the study
\cite{yhat} is an enhancement of the viscosity $\eta$ 
%($\eta \propto \tau$)
of 
the bacterial bath. This enhanced viscosity is proportional to the large
time scale $\tau_c$ emerging in the problem. In this paper we show that 
this {\it crossover time scale $\tau_c$} is connected to a 
{\it viscous to an elastic
transition in a bacterial bath.} \footnote{We emphasize that the main 
aspect of a bacterial bath that we are interested
in is the presence of correlated structures in the form of swirls. Such 
structures give rise to a large spatial scale in the problem. 
This in turn leads to a large crossover time scale $\tau_c$ which is 
of the order of the lifetime of a swirl.} 
%This suggests a viscoelastic model.
Let us consider the following model which interpolates 
between a purely viscous response and a 
purely elastic one for a liquid of viscosity $\eta$ and 
high frequency rigidity modulus $G$. 
\begin{equation}
(\frac{1}{\eta}+\frac{1}{G}\frac{\partial}{\partial t})\sigma_{xz} 
= \frac{\partial v_x}{\partial z}
+\frac{\partial v_z}{\partial x}.
\label{vis}
\end{equation}
Here $\sigma_{xz}$ is the $xz$ component of the shearing stress tensor 
corresponding to the strain rate $\frac{\partial v_x}{\partial z}
+\frac{\partial v_z}{\partial x}$.
On taking the Fourier-Laplace transform of Eq.(\ref{vis}) one notices that the 
viscoelastic approximation corresponds to replacing the zero frequency 
shear viscosity $\eta$ by a frequency dependent shear viscosity:
\begin{equation}
\tilde{\eta}(\omega) 
= \frac{G}{(-i\omega +\frac{G}{\eta})}.
\label{vislap}
\end{equation}
Defining $\tau_c =\frac{\eta}{G}$ to be the relaxation time, we can
rewrite Eq. (\ref{vislap}) as  
\begin{equation}
\tilde{\eta}(\omega)
= \frac{G}{(-i\omega +\frac{1}{\tau_c})}.
\label{vislap2}
\end{equation}
%$$\eta = \frac{k \tau}2{6\pi a (1+9a/16d)$$
Therefore on long time scales for which $\omega \tau_c \ll 1$ one 
expects the flow to be viscosity ($\eta=G\tau_c$) dominated. 
As one starts probing shorter
time scales, one expects a transition from 
a viscous response to an elastic one to take place for $\omega \tau_c \geq 1$.
Since for a bacterial bath $\tau_c \approx 1 sec$ we arrive at the following 
important quantitative point. {\it Unlike a 
colloidal bath where the onset of propagating shear waves takes
place on time scales $\approx 10^{-5} sec$, in a bacterial bath 
we predict that this would happen on time scales 
$\approx 1 sec$}. Thus, if a bacterial bath is perturbed at a frequency 
of $\approx 1Hz$ one would notice a transition from a viscous relaxing
behavior to an {\it elastic propagating one}. 
Piezoelectric bimorphs can be used
to produce and detect controlled shear motions at a 
frequency $\approx 1 Hz$ \cite{science} in a bacterial bath.
In contrast, in order to see
the propagation of shear waves in a  
colloidal liquid one needs to perturb it at a high 
frequency $\approx 10^5 Hz$. A bacterial bath thus provides us with 
a ``scaled up'' non-equilibrium statistical mechanical system.
This is the key qualitatively and quantitatively interesting 
observation made in this paper.
We will analyze the emergence of 
shear waves in some more detail below. 
 
An analysis based on memory function formalism 
can be used to obtain a statistical 
mechanical description of the transverse current fluctuations in a 
bacterial bath.
In this formalism the spectrum  $\tilde{C}_{T}(k,\omega)$ of transverse 
current fluctuations ( i.e. the
Fourier-Laplace transform of the transverse current autocorrelation
function $C_{T}(k,t)=<j^{T}_{\vec{k}}(t)j^{T}_{-\vec{k}}(0)>$ )
takes the form:
\begin{equation}
\tilde{C}_{T}(k,\omega)
=\frac{C_T(k,t=0)}{[-i\omega+\frac{k^2}{\rho}
\tilde{M}_{T}(k,\omega)]} .
\label{transc}
\end{equation}
where $\tilde{M}_{T}(k,\omega)$ is the pertinent memory function 
which converges to the hydrodynamic viscosity $\eta$ in the long 
wavelength
and long time limit. Modelling the memory function in the time domain 
as ${M}_{T}(k,t)= {M}_{T}(k,t=0)e^{-t/\tau_c(k)}$ we arrive at the 
following expression for the real part ${C}_{T}(k,\omega)$ of the 
spectrum  $\tilde{C}_{T}(k,\omega)$ of transverse
current fluctuations: 
\begin{equation}
{C}_{T}(k,\omega)
=\frac{{\omega_0}^2\omega_{1t}^2\tau_c(k)}{\pi [\omega^2+\tau_c^2(k)
(\omega_{1t}^2 - \omega^2)^2]} .
\label{transcurr}
\end{equation}
where $\tau_c(k)$ is the wave vector dependent lifetime of correlated 
structures in the bacterial bath; $\omega_0^2=C_{T}(k,t=0)$
and $\omega_{1t}^2=\frac{k^2 G}{\rho}$, with $\rho$ the mass 
density of a bacterial 
swirl. 
We can now establish the criterion for the existence of propagating 
transverse modes.
From Eq.(\ref{transcurr}) it is clear that the condition for 
${C}_{T}(k,\omega)$
to have a peak at a nonzero frequency, i.e. the 
threshold condition for shear waves to 
propagate at a finite speed is $\omega_{1t}^2 \tau_c^2(k) = \frac{1}{2}$.
The corresponding value for the critical wavevector is 
$k_c=\frac{2 \pi}{l_c}=\frac{1}{\tau_c}\sqrt{\frac{\rho}{2 G}}$ 
which corresponds to 
a length scale of the order of the spatial scale $l_c$ of a bacterial swirl. 
Thus, as one starts sweeping over wavevectors, the form of the spectrum
${C}_{T}(k,\omega)$ changes from a profile peaked around 
$\omega \tau_c=0$ for $kl_c<<1$  to one peaked around a finite nonzero 
frequency $\omega \tau_c \approx 1$ for $kl_c\approx 1$ (See Fig. $1$).
\begin{figure}[htbp]
\centering{\resizebox{4cm}{!}{\includegraphics{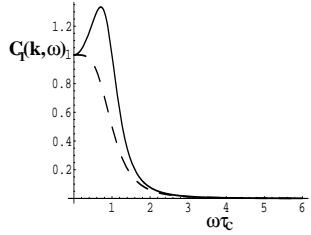}}}
%\centering{\frame{\rule{2cm}{2cm}}}
\caption{Schematic representation of spectrum of transverse current 
autocorrelation function
$C_T(k,\omega)$ as a function of $\omega\tau_c$. The figure shows
transition from a profile centered around  $\omega\tau_c=0$ (dashed 
curve)
for $kl_c<<1$
to a profile with a side peak at $\omega\tau_c =1$ for $kl_c=1$
illustrating
the phenomenon of onset of shear waves.} 
\end{figure}

Using $\tau_c \approx 2 sec$, $l_c \approx 10 \mu m $ and an appropriate
value of $\rho$ based on 
experimental observations \cite{libchaber} one can thus calculate the 
high frequency rigidity modulus $G$ for a bacterial bath. 

\section{Discussion}
There have been some interesting 
studies of Brownian motion of spherical 
probes in wormlike micelles\cite{mason} where they study the role of 
polymer solutions in determining diffusion behavior of probes.
The focus of the work presented here is quite different from that. 
We focus on a super-diffusive to diffusive behavior of the type that is 
observed, for 
instance, in a simple liquid where a ballistic to diffusive transition 
takes place at a typical relaxation time scale \cite{hansen,Dhont}.
This leads to an interesting viscoelastic effect. 
To summarize, we predict an {\it onset of shear 
waves} in a bacterial bath at $\approx 1 Hz$
as a result of structural ordering on {\it larger length scales} 
compared to 
that in a dense viscous colloidal liquid. This is a 
consequence of a large spatial scale arising from the  
swirl size of the bacteria forming the bath. 
This enables us to view a bacterial bath as 
a `scaled up' dense viscous liquid. 
Furthermore, our work points to an {\it independent test} of 
structural ordering in a bacterial bath. 
We expect this work to generate interest in testing the 
predictions through computer simulations and experiments in bacterial 
baths. 
\section*{Acknowledgements}
It is a pleasure to thank E. G. D. Cohen 
and M. C. Marchetti 
for insightful comments on propagation effects in dense liquids.  
I also thank  
V. Shenoy, Abhishek Dhar
and G. V. Shivashankar 
for discussions.

\end{document}